# Exchange Bias Induced by the $Fe_3O_4$ Verwey transition

J. de la Venta[1,*], M. Erekhinsky[1], Siming Wang[1,2], K. G. West[1], R. Morales[3,4] and Ivan K. Schuller[1]

*1.- Department of Physics and Center for Advanced Nanoscience, University of California San Diego, La Jolla, CA 92093 USA*
*2.- Materials Science and Engineering Program, University of California San Diego, La Jolla, CA 92093 USA*
*3.- Dpto. de Química-Física, Universidad del País Vasco, 48940 Leioa, Spain.*
*4.- IKERBASQUE, Basque Foundation for Science, 48011 Bilbao, Spain.*

*) Corresponding author E-mail: jdelaventa@physics.ucsd.edu

Abstract

We present a study of the exchange bias in different configurations of $V_2O_3$ thin films with ferromagnetic layers. The exchange bias is accompanied by a large vertical shift in the magnetization. These effects are only observed when $V_2O_3$ is grown on top of $Ni_{80}Fe_{20}$ permalloy. The magnitude of the vertical shift is as large as 60% of the total magnetization which has never been reported in any system. X-Ray diffraction studies show that the growth conditions promote the formation of a ferrimagnetic $Fe_3O_4$ interlayer. The change in the easy magnetization axis of $Fe_3O_4$ across the Verwey transition at 120 K is correlated with the appearance of exchange bias and vertical shift in magnetization. Both phenomena disappear above 120 K, indicating for the first time a direct relationship between the magnetic signature of the Verwey transition and exchange bias.

I. INTRODUCTION

Exchange bias (EB)[1,2], the horizontal shift of the hysteresis loop in ferromagnetic/ antiferromagnetic (or ferrimagnetic) systems, is one of the most studied effects in thin film magnetism and is the basis for several applications [3,4]. Most studies of EB in thin films were done by growing ferromagnetic (FM) layers on top of antiferromagnetic (AFM) oxides. These oxides exhibit unique properties including collective phenomena such as magnetism, ferroelectricity, superconductivity, metal-insulator transitions, enhanced photoconductivity and electron transfer [5,6,7,8,9,10]. These materials hold great potential for revolutionary developments which may arise from the manipulation and control of novel emergent properties which are not present in ordinary semiconductors or metals. Interesting possibilities arise when FM thin films are layered with transition metal oxides that undergo a structural transition where there is a change in resistivity and magnetic properties. This transition provides an external tuning capability of the properties by changing the temperature. It affords a unique opportunity to study EB not only when the materials are cooled through the Neel temperature but also when their structure and resistivity change across the transition.

A particularly interesting class of oxides is the vanadium oxides $VO_x$ [11,12,13] system. Among these phases, vanadium sesquioxide, $V_2O_3$, undergoes a first-order metal-insulator transition at T=160 K from a low temperature antiferromagnetic insulating phase to a high temperature paramagnetic metallic phase [14,15,16]. The crystal structure changes from monoclinic in the insulating phase to rhombohedral symmetry in the metallic phase. Electronic and structural properties of $V_2O_3$ thin films epitaxially grown on different sapphire ($\alpha$-$Al_2O_3$) planes have been extensively studied [17,18]. Although the phase diagram [16] suggests that a reduction in the metal-insulator transition may occur for some strain values, Allimi et al. [18] showed how the transition temperatures of $V_2O_3$ films grown on (11$\bar{2}$0) a- and (0001) c-plane sapphire were enhanced in both films.

The aim of this work is to study EB in $V_2O_3$ with different ferromagnetic thin films. For this purpose, different combinations of ferromagnetic (FM) and antiferromagnetic ($V_2O_3$) layers have been grown on (1$\bar{1}$02) r-plane sapphire. EB has previously been observed in bilayers of $V_2O_3$ grown on (11$\bar{2}$0) a- plane sapphire substrates with Co on top[19]. However the EB is absent when the top layer is Fe, which was attributed to the presence of a magnetically "dead" Fe layer at the interface. Additional investigations of Co, Ni and Fe on top of $V_2O_3$ grown on (0001) c-plane and (11$\bar{2}$0) a- plane substrates[20] found EB for all of the combinations except for Fe on $V_2O_3$ (11$\bar{2}$0).

Here we report the observation of substantial EB in permalloy (Py)/$V_2O_3$ bilayers with a large vertical shift in magnetization never observed before in any system. From a series of well-defined experiments: i) X-ray diffraction, ii) Magnetization and magnetotransport iii) $Cl_2$ plasma etch, iv) annealing, v) trilayer and vi) (Ni)/$V_2O_3$ and (Py)/$V_2O_3$ bilayers we show that the formation of a $Fe_3O_4$ interfacial layer produces both effects. The results imply that the change in the easy magnetization axis occurring at the $Fe_3O_4$ Verwey transition is responsible for the EB and the vertical shift in the FM magnetization.

## II. EXPERIMENTAL

The samples were prepared in a high vacuum sputtering deposition system with a base pressure of $1\times10^{-7}$ Torr using ultrahigh purity (UHP) argon sputtering gas. The pressure during the deposition was $4\times10^{-3}$ Torr. Bilayers and trilayers were deposited by RF sputtering of $Ni_{80}Fe_{20}$, Ni, and $V_2O_3$ targets on r-plane sapphire ($1\bar{1}02$) substrates. The different configurations are shown in figure 1. The RF magnetron power was kept at 100 W. In the case of FM/$V_2O_3$ bilayer (figures 1a and 1b), the ferromagnetic layer was deposited at room temperature. Next, the temperature was increased to 750°C at 7° C/min and the $V_2O_3$ layer was deposited. Finally the sample was cooled to room temperature at 7°C/min. In the case of $V_2O_3$/FM bilayer (figure 1c), the $V_2O_3$ layer was deposited at 750°C. After cooling the sample to RT, the FM layer was deposited. Samples with different thicknesses for FM layers (20 nm to 60 nm) and $V_2O_3$ (25 nm to 100 nm) were fabricated. For the FM/$V_2O_3$/FM trilayers (figure 1d) the two bottom layers were deposited as explained for FM/$V_2O_3$ bilayers and the top FM was deposited using two different conditions: at room temperature and at 750°C.

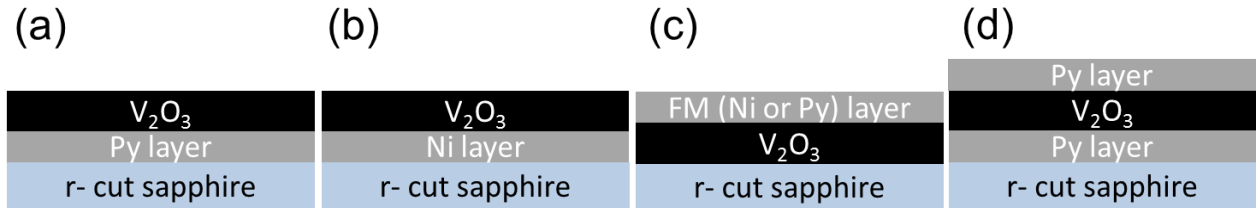

FIG. 1 (color online). Schematic model of different samples studied. (a) and (b) The Py or Ni was deposited at room temperature. The $V_2O_3$ layer was deposited later at 750°C after a 7 K/min heating ramp. (c) The $V_2O_3$ layer was deposited at 750°C and the top FM at room temperature (d) Trilayer used for MOKE measurements.

X-ray diffraction was performed with a Bruker D8 Discovery rotating anode X-ray diffractometer using Cu K$\alpha$ radiation.

Magnetic characterization was performed using a superconducting quantum interference device magnetometer (SQUID) and the magneto-optical Kerr effect (MOKE). While MOKE is only sensitive to the surface magnetization (the penetration length of the light is less than 50 nm for the FM layer) the SQUID measures the full magnetic moment of the sample. The samples were cooled from 250 K to 10 K in different cooling fields. Hysteresis loops were measured at various temperatures after each field cooling.

In addition, the magnetoresistance of Py/$V_2O_3$ bilayers were measured on 40 µm wide bars etched out of the whole film using standard photolithographic techniques. Since the resistivity of $V_2O_3$, even in the metallic phase, is much higher than the resistivity of FM layer, the majority of the current in the film flowed through the ferromagnetic metal layers. Therefore the anisotropic magnetoresistance (AMR) reflects the switching fields of the ferromagnetic layer.

### III. RESULTS AND DISCUSSION.

#### A. X-ray Diffraction.

The X-ray diffraction of FM/$V_2O_3$ bilayers (figure 1a and 1b) shows the following diffraction peaks (figure 2): Py or Ni in the (111) direction; $V_2O_3$ in the (001) direction. The Py/$V_2O_3$ bilayers (figure 1a) exhibit a series of unexpected (*with instrumentally-limited widths*) diffraction peaks, marked as dots in figure 2. Two possible candidates match the X-ray diffraction peaks for this unexpected phase: $V_3O_4$ (111) or magnetite $Fe_3O_4$ (111), both with the same cubic *F d -3 m* structure. The peak positions for (111) family planes of bulk $Fe_3O_4$ (or $V_3O_4$) are marked as lines at the top of figure 2. These unexpected diffraction maxima did not appear when the $V_2O_3$ thin film is deposited on top of a Ni layer. A literature review indicates that the existence of the $V_3O_4$ phase is uncertain. This was claimed only twice in the literature; ultra-thin, less than 1 monolayer [21] or $V_3O_4$ spinel phase generated from $V_2O_5$ by compressive shock waves stabilized by an Fe admixture. [22]

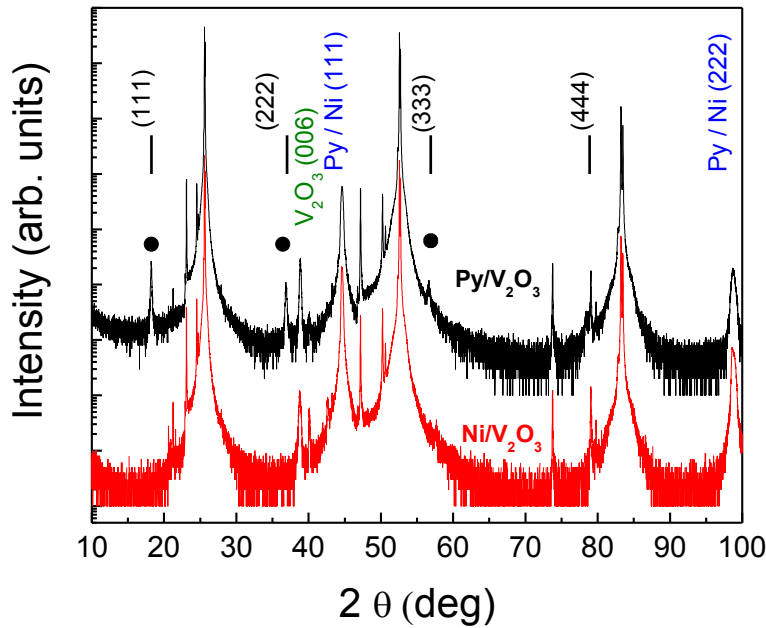

FIG. 2 (color online). X-ray diffraction patterns for two bilayers (top black) Py(30 nm)/$V_2O_3$ (100 nm) ; (bottom red) Ni (30 nm)/$V_2O_3$ (100 nm). Diffraction maxima from Py or Ni (111) and from $V_2O_3$ (006) are marked in the figure; (●) diffraction maxima from an unexpected phase. Peak positions from $Fe_3O_4$ or $V_3O_4$(111) planes are marked at the top. All other diffraction maxima not marked in the figure are due to the r-cut sapphire substrate. *Note that the (444) peak overlaps with one of the substrate peaks*. See supplementary material for the substrate X-ray diffraction pattern.

B. Magnetization and Magnetotransport

Surprisingly only the Py/$V_2O_3$ bilayers (figure 1a) showed EB field with a large vertical shift in the magnetization ($M_{Shift}$), figure 3. Both the EB and $M_{Shift}$ vanish above 120 K. No EB was found in Ni/$V_2O_3$ bilayers (figure 1b). In addition, bilayers with the reverse configuration, i. e. $V_2O_3$ at the bottom and FM (Py or Ni) on top (figure 1c) show no EB. Hence, the presence of EB is related with an interaction between the bottom Py layer and the top $V_2O_3$ thin film.

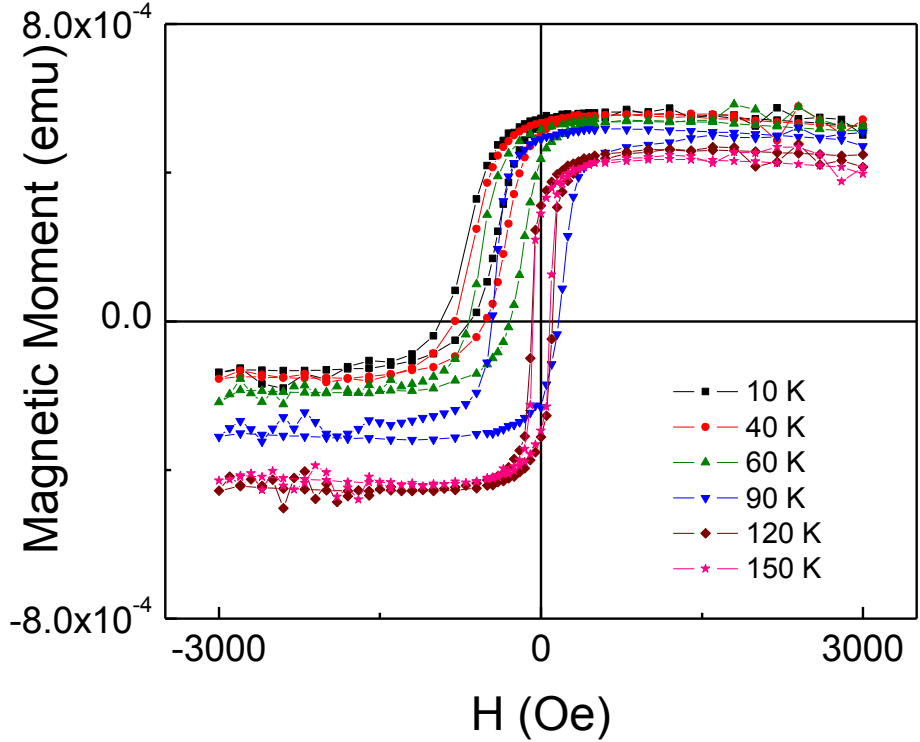

FIG. 3 (color online). Hysteresis loops at different temperatures in Py (30 nm)/$V_2O_3$ (100 nm) bilayers. The sample was cooled from room temperature to 10 K with a field of 1000 Oe.

The most striking effect is the large vertical shift, up to 60% of the total magnetization at 10 K, for Py (30 nm)/$V_2O_3$ (100 nm), black triangles in figure 4. Such large magnetization shifts were never reported in any system. In $FeF_2$-Fe bilayers, the shift of the hysteresis loop is about 1% of the total magnetization.[23] X-ray magnetic circular dichroism shows, a small vertical offset in Co/$Ir_{20}Mn_{80}$ bilayers, indicating that only a 7 % of the total uncompensated moments were pinned.[24] The large vertical shift in the Py/$V_2O_3$ bilayers indicates the presence of a large number of pinned spins. The influence of the field cooling has been investigated at fields of up to 5 T. In all cases, the shift of the hysteresis loop is "upwards" for positive cooling fields. No dependence of the $M_{Shift}$ with the magnitude of the cooling field or indication of positive EB was found. These findings imply that the coupling between the different layers is ferromagnetic.[23] Samples with 25 nm of $V_2O_3$ show no EB or $M_{Shift}$, red squares in figure 4.

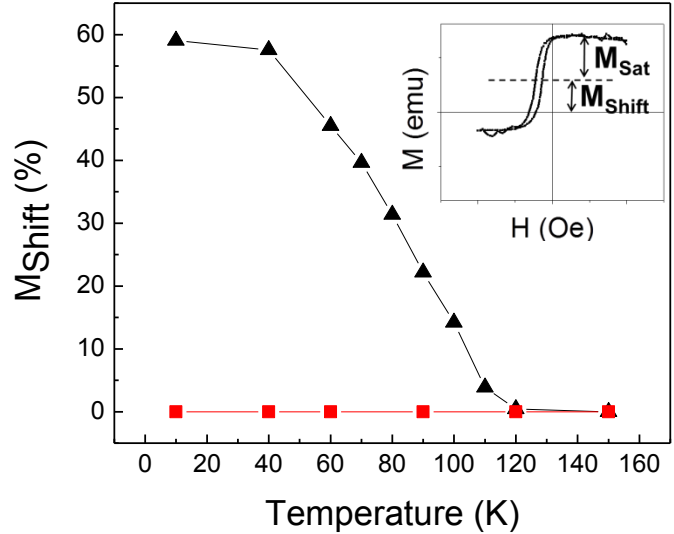

FIG. 4 (color online). Vertical shift in the magnetization ($M_{Shift}$) values as a function of the temperature for bilayers of different thicknesses, Py (30 nm) /$V_2O_3$(y): ▲ (100 nm); ■ (25 nm). $M_{Shift}$ is absent for the 25 nm $V_2O_3$ film. The cooling field value was 1000 Oe. (Inset) $M_{Shift}$ (%) has been calculated from the shift of the loop with respect to the abscissa axe and normalized to the values of the saturation $M_{Sat}$ at positive fields.

Magnetoresistance (MR) measurement of Py (30 nm)/$V_2O_3$ (50 nm) bilayers confirmed the presence of EB in samples containing Py/$V_2O_3$ interfaces. After the sample was cooled to 4.2 K in 1000 Oe positive magnetic field the magnetoresistance was measured at different monotonically increasing temperatures. Figure 5 shows the H field dependence of resistance at different temperatures. For each temperature there are two curves corresponding to different magnetic field sweep directions.

Below 100K the MRs are centered around a non-zero field, indicating the presence of EB. As the magnetic field is swept from a positive to negative (or vice versa) then the distance between minima in the MR indicates the coercivity of the FM layer. At 5 K and 20 K the curves match giving the same values of EB. At 100K, the bilayer MR is almost symmetric around zero field and the midpoint of the MR gradually shifts towards negative field. This indicates that the blocking temperature obtained from the MR is around 100 K.

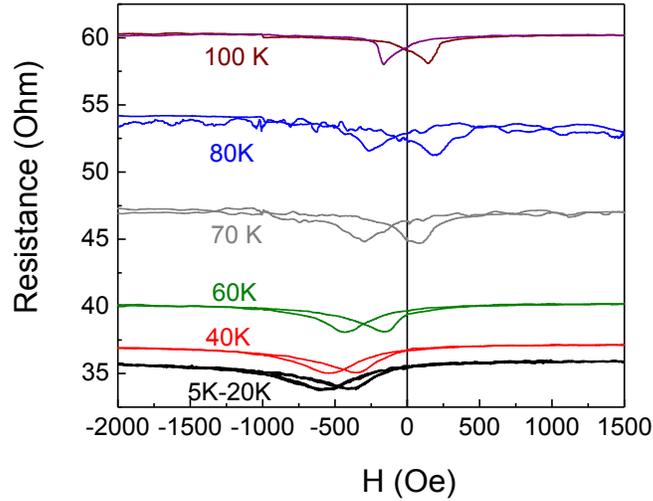

FIG. 5 (color online). Magnetoresistance measurements at different temperatures in Py (30 nm)/$V_2O_3$ (50 nm). The sample was cooled from 300 K to 4.2 K with an applied field of 1000 Oe.

The temperature dependence of the exchange bias ($H_{EB}$) and coercivity ($H_C$) for 30 nm of Py and different $V_2O_3$ thicknesses are plotted in figure 6. The $H_{EB}$ and $H_C$ extracted from SQUID and MR measurements are in good quantitative agreement for Py (30 nm)/$V_2O_3$ (50 nm), solid and empty circles in figure 6. The $H_{EB}$ and $H_C$ are slightly larger for the bilayer with a thicker $V_2O_3$ film (100 nm), indicating that it is not only due to a surface effect. [25] The blocking temperature, i. e. the temperature at which the EB disappears is 120 K in all cases. When the thickness of the $V_2O_3$ layer is 25 nm neither EB nor $M_{Shift}$ have been observed, red squares in figures 6 and 4 respectively.

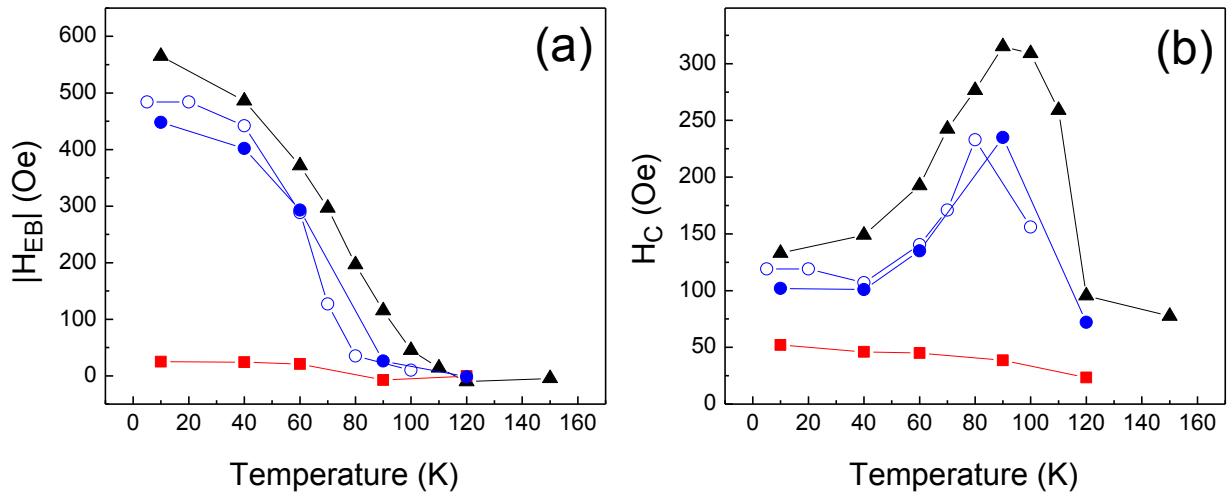

FIG. 6 (color online). (a) Dependence of the exchange bias field with temperature; (b) dependence of the coercive field with temperature. (—▲—) Py (30 nm)/$V_2O_3$ (100 nm) bilayers measured with SQUID; (—●—) Py (30 nm)/$V_2O_3$ (50 nm) measured with SQUID; (—○—) Py (30 nm)/$V_2O_3$ (50 nm) from magnetoresistance measurements; (—■—) Py (30 nm)/$V_2O_3$ (25 nm) measured with SQUID.

C. $Cl_2$ plasma etch.

To investigate the cause of the EB and to clarify the origin and composition of the unexpected phase found from X-ray diffraction, (figure 2) one of the samples was subjected to reactive ion etching. The $Cl_2$ plasma quickly etches vanadium and its oxides but not Py. The $V_2O_3$ (006) X-ray diffraction peaks disappear after etching, figure 7, whereas the diffraction maxima arising from the new phase remain. After etching the EB was still present with $H_{EB}$ and $M_{Shift}$ as in the virgin samples. This indicates that the unexpected phase is $Fe_3O_4$ rather than $V_3O_4$.

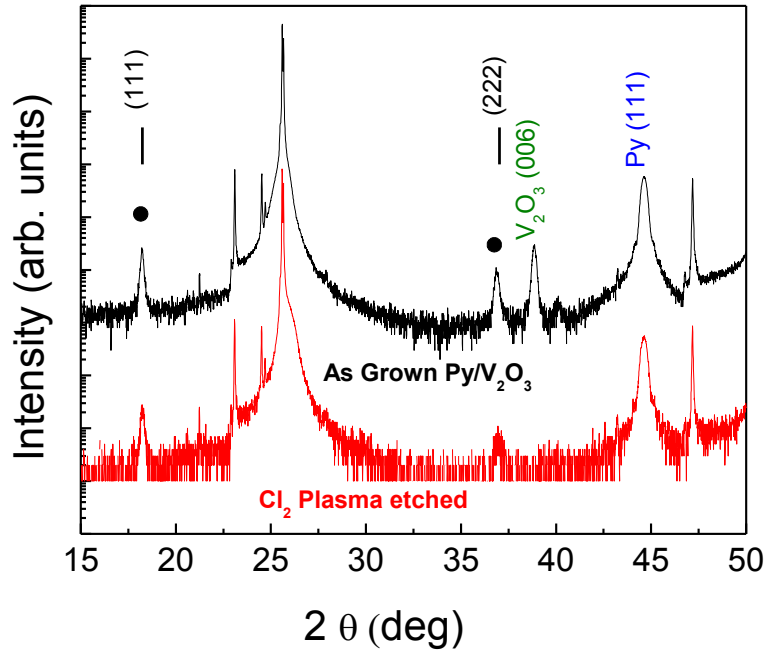

FIG. 7 (color online). X-ray diffraction patterns for a bilayer Py(30 nm)/ $V_2O_3$ (100 nm) (top black) as grown; (bottom red) after etching. Diffraction maxima from $V_2O_3$ and Py are labeled in the figure; (●) diffraction maxima from an unexpected phase. Peak positions from $Fe_3O_4$ (111) planes are marked at the top. All other diffraction maxima not marked in the figure are due to the r-cut sapphire substrate.

D. Annealing.

Figure 8a shows the diffraction maxima at 36.86° from $Fe_3O_4$ (222) and at 38.84° from $V_2O_3$ (006). Figure 8b shows the peaks at 44.6° from Py (111). Bilayers with 30 nm Py and different thicknesses of $V_2O_3$ are included. For better visualization the X-ray data have been smoothed and XRD intensity in figure 8a is plotted in linear scale whereas in figure 8b the vertical scale is logarithmic. (For raw data is See Supplemental Material).

The Py (30 nm)/$V_2O_3$ (25 nm) bilayer which did not show EB or $M_{Shift}$ (red squares in figures 4 and 6) shows $V_2O_3$ (006) XRD peaks but not those from the unexpected phase, figure 8a. Hereafter we refer to this sample as the unreacted bilayer. As an additional test this bilayers was subjected to thermal treatment. After depositing the $V_2O_3$ layer on top of Py the temperature was kept at 750°C for 18min 45 sec (25 min total at 750°C). This is the same deposition time for 100nm $V_2O_3$ at which the heater was at 750°C for 25min. After this annealing, both the new diffraction maxima (figure 8a) and EB appeared. The short deposition time (4 min 15 sec) for a 25 nm $V_2O_3$ sample is not enough to produce the interface reaction giving rise to the $Fe_3O_4$ interfacial layer. However, stopping the deposition but maintaining the temperature provides the conditions for the $Fe_3O_4$ layer formation. The $Fe_3O_4$ (222) peaks in the samples with 50 nm and 100 nm $V_2O_3$ have the same intensity and sharpness. This is an additional indication that the thickness of the $Fe_3O_4$ layer is similar in both cases. As expected the peak from $V_2O_3$ (006) has a higher intensity in 100 nm of $V_2O_3$. The intensity from the 38.84° peak for the unreacted bilayer indicates that the $V_2O_3$ thickness is between those bilayers with 50 and 100 nm of $V_2O_3$.

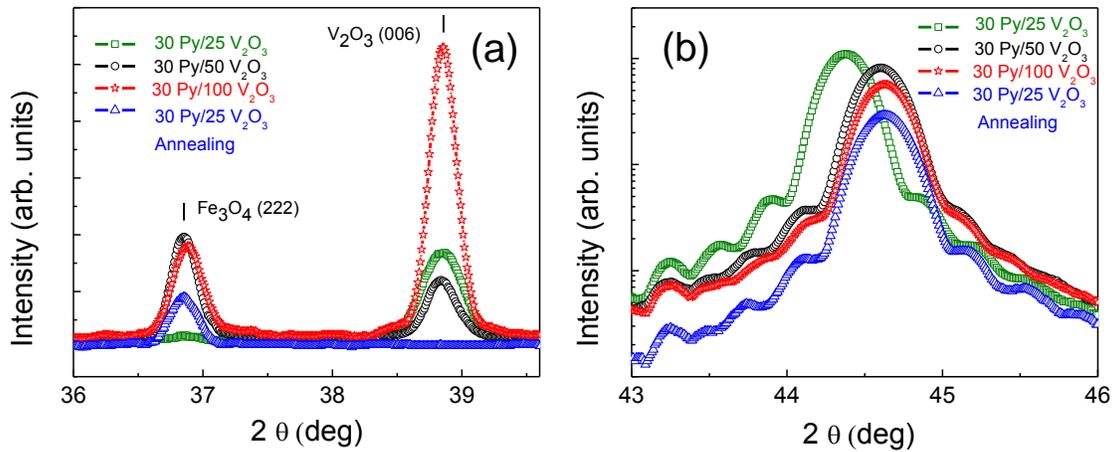

FIG. 8 (color online). X-ray diffraction patterns focused around the maxima from: (a) $Fe_3O_4$ (222) and $V_2O_3$ (006); (b) Py (111). –□– 30 nm Py/25nm $V_2O_3$ ;–○–30 nm Py/50 nm$V_2O_3$ ;–☆–30 nm Py/100 nm$V_2O_3$ ;–△–30 nm Py/25 nm$V_2O_3$ annealed. The $Fe_3O_4$ (222) is located at 36.86°; the $V_2O_3$ (006) at 38.84°; the Py peaks shifts from 44.38° from to 44.6° and 44.63°.

The peaks close to 44.5°, figure 8b, provide further clues regarding the interface reaction of the Py film. In all the bilayers the peak position shifts (gradually) towards higher angles (44.6°, 44.63°) away from the unreacted bilayer (44.38°). The corresponding lattice parameters are 3.5124 Å, 3.5147 Å and 3.5312 Å. The latter is closer to the Py bulk value 3.5507 Å while the first are closer to the bulk Ni value 3.5238 Å. The finite size Laue oscillations around the Bragg peak for the unreacted sample are more intense and clearer than the ones of the reacted samples indicating that the interfaces are sharper. The finite size oscillations imply that the Py thickness is 28 nm for the unreacted bilayer as opposed to 23-24 nm for the rest of the samples. Thus the Py thickness is reduced approximately 5 nm by the reaction. Thus the unreacted sample shows a smoother Py surface and a lattice parameters closer to bulk Py. When reaction to form $Fe_3O_4$ takes place the surfaces become rougher and the lattice parameter changes closer to bulk Ni. All the above suggests Fe migration from Py to form the $Fe_3O_4$ layer.

E. Trilayers

MOKE measurements on Py (30 nm)/ $V_2O_3$ (100 nm) /Py (30 nm) trilayers (figure 1d) grown on double side polished sapphire are shown in figure 9. MOKE measurements are only sensitive to surface magnetization since the light penetration length in the FM layer is about 50 nm. Note that MOKE is not a measure of the absolute M value so it can only confirm the presence of EB field but not the vertical shift in the magnetic moment. In agreement with the previous ideas EB is only found in the bottom layer, figure 9. As a further check, EB is absent for either the top Py layer deposited at room temperature or even at 750° C. Thus, the $Fe_3O_4$ (111) interface is created when the $V_2O_3$ is deposited on top of the Py.

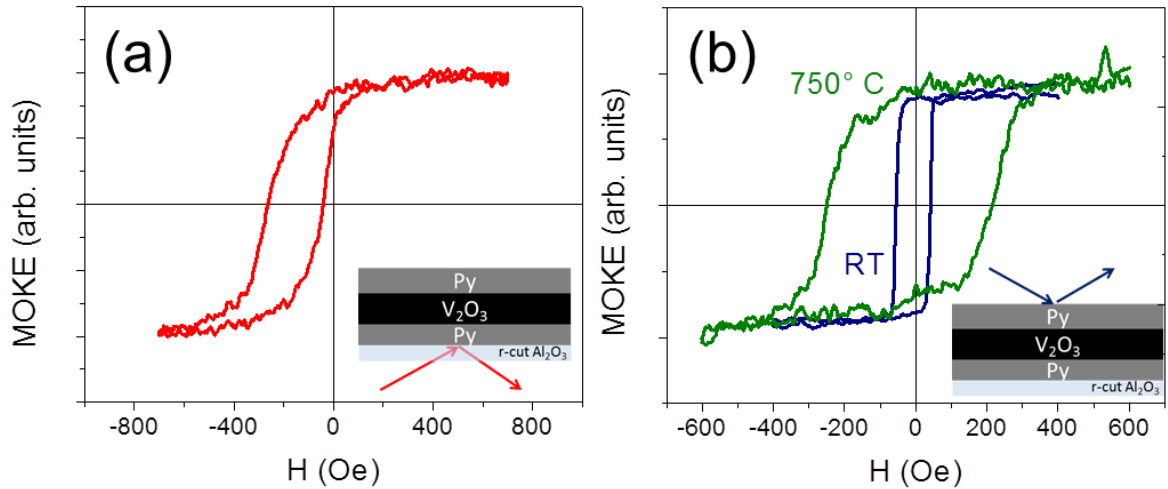

FIG. 9 (color online). MOKE hysteresis loops at 10 K after cooling the sample with a field of 1000 Oe. (a) MOKE probing bottom Py layer. (b) MOKE probing top Py layer deposited at room temperature and at 750° C. Insets: MOKE probing each FM individual layer.

Although the top Py layer exhibits no EB independently of the deposition temperature its properties are different when deposited at room temperature than 750 °C. For room temperature deposition the top layer has the characteristic metallic shiny aspect of metals. Whereas at higher temperature the surface becomes opaque thus the reflected laser intensity is somewhat reduced and the MOKE signal is nosier, figure 9b. Moreover the coercive field of this top layer is smaller for room temperature deposition, (figure 9b) which also implies smoother interfaces.The XRD data (see supplementary material) shows that the $Fe_3O_4$ and $V_2O_3$ interlayers are formed either when the top layer is deposited at room temperature or at 750° C.

F. Discussion.

All the extensive experimental evidence discussed above, shows that an unexpected interfacial phase is formed by interfacial diffusion or reaction. There are three timescales which are relevant: A) Short deposition times, without $Fe_3O_4$ formation; the Py preserves its bulk-like characteristics, the surfaces are smoother (figure 8b) and the $V_2O_3$ deposited on top of Py "survives". This latter point explains also the reason for the $V_2O_3$ peak intensities being larger for 30 nm $V_2O_3$ than 50 nm. B) Intermediate deposition times, causing interface reaction between the Py and $V_2O_3$ to form the $Fe_3O_4$ layer. The Py lattice parameter changes closer to bulk Ni and the interfaces become rougher. A similar process to medium deposition times occurs when annealing the samples for longer periods after deposition. In this case the 25 nm $V_2O_3$ layer is destroyed during the process (figure 8a) giving rise to the formation of $Fe_3O_4$ C) Longer deposition time, after the reaction takes place the $Fe_3O_4$ and Py thicknesses and their lattice parameters are stable and $V_2O_3$ continues growing on the newly formed $Fe_3O_4$.

All the magnetic data is consistent with the EB and the $M_{Shift}$ being caused by the magnetic interaction between the Py and an interfacial $Fe_3O_4$. Further evidence is provided by the fact that when $V_2O_3$ is deposited on a Ni layer, neither EB nor the presence of the new $Fe_3O_4$ phase were found.

$Fe_3O_4$ is a ferrimagnetic material with a Curie temperature of 825 K which changes lattice symmetry and exhibits a metal-insulator transition at ~ 120 K ($T_v$), the Verwey transition. [26,27] The change in crystal structure is accompanied by magnetization changes. The blocking temperature for EB found here coincides with the $Fe_3O_4$ Verwey transition and is substantially lower than the 160K AFM transition of the $V_2O_3$ as shown in figure 6a. This again indicates that the EB and $M_{Shift}$ vanishing above 120 K should be attributed to the $Fe_3O_4$ Verwey transition. Earlier measurements using magnetite $Fe_3O_4$ showed EB which vanishes at 200 K and 275 K, due to antiphase boundaries [28] and a spin-glass-like phase [29], respectively. FM/$Fe_3O_4$ bilayers showed EB, [30,31] with smaller values of $H_{EB}$ and no $M_{Shift}$ without indications of a blocking temperature. On the other hand, the EB found in AFM/$Fe_3O_4$ bilayers is attributed to presence of the AFM phase [32,33]. To the best of our knowledge this is the first direct observation of EB which is correlated with the appearance of the Verwey transition in $Fe_3O_4$.

Although the Verwey transition is known for more than 70 years, its origin is still controversial. The consensus is that the magnetic signature of the Verwey transition manifests as a shift of the easy magnetization axis from the [111] to the [100] direction. [34,35,36,37] On the other hand, some claim that the two transitions are uncoupled and refer to the slightly higher (130 K) magnetic transition temperature as the "isotropic point" of $Fe_3O_4$. [38,39] A recent report shows that the small $Fe_3O_4$ lattice distortion in the c direction may account for the change in magnetic easy-axis from cubic [111] to [001] at temperatures less than $T_V$. [40] This debate is beyond the scope of the present paper. Since here the $Fe_3O_4$ grows in the (111) orientation, above 120 K the easy magnetization [111] axis is perpendicular to the FM easy axis. Thus above 120 K there is no magnetic coupling between the two layers with the consequent absence of EB and the $M_{Shift}$, figure 10a. Below 120 K, the easy axis of $Fe_3O_4$ changes to the [100] direction at 45° with the Py as shown in the inset of figure 10b. Therefore the non zero in plane magnetization component of $Fe_3O_4$ coincides with the easy in plane axis of the FM layer, giving rise to the EB and the $M_{Shift}$, figure 10b. The bilayers studied here are textured as proven clearly by the X-ray diffraction data and the cartoons just reflect the magnetic orientation in textured samples.

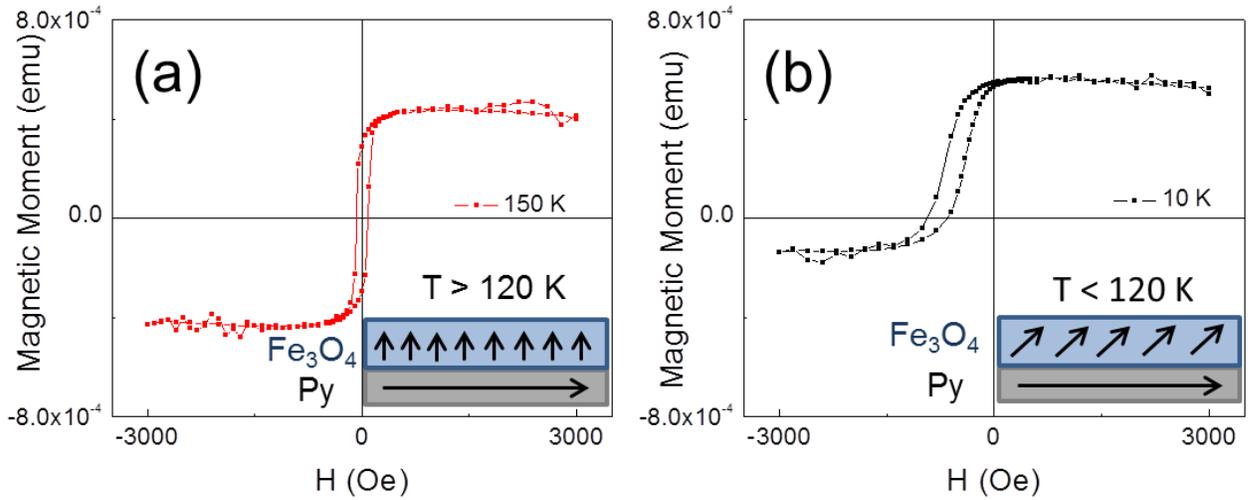

FIG. 10 (color online). Hysteresis loops in Py (30 nm)/$V_2O_3$ (100 nm) bilayers after field cooling at: (a) 150 K and (b) 10 K. (Insets) Simplified magnetization easy axes configuration of the Py/$Fe_3O_4$ bilayer: (a) [111] above the transition (b) [100] and below the transition.

## IV.   CONCLUSION

We studied the magnetic properties of FM/$V_2O_3$ bilayers grown on r-cut sapphire in a variety of configurations. The exchange bias in Py/$V_2O_3$ is accompanied by a large (60%) positive vertical shift in the magnetization, indicating the presence of a substantial number of pinned spins. A series of experiments show these effects are related to the formation of a $Fe_3O_4$ ferrimagnetic interfacial layer. The magnetic signature of the $Fe_3O_4$ Verwey transition at 120 K coincides with EB blocking temperature. The large EB and the $M_{Shift}$ are produced by an interfacial ferromagnetic coupling due to changes in the $Fe_3O_4$ magnetic anisotropy. This is the first observation in which the magnetic signature of the Verwey transition is directly responsible for the presence of exchange bias.

## AKNOWLEDGEMENTS

This work has been supported by the US Department of Energy, Office of Basic Energy Sciences, Division of Materials Sciences under Award FG03-87ER-45332. J. de la Venta acknowledges the support of a postdoctoral fellowship from *Ministerio de Educacion* of Spain. R. M. acknowledges support from Spanish grant MICINN FIS2008-06249 and funding from IKERBASQUE Basque Foundation for Science